\documentclass{mem}
\usepackage{natbib}\usepackage{txfonts}\usepackage{balance}
\usepackage{graphicx}
\usepackage[a4paper]{hyperref}
\idline{00}{000}
\input epsf.sty
\begin{document}
\def\teff{$T\rm_{eff }$}
\def\kms{$\mathrm {km s}^{-1}$}

\title{
 Galactic Globular Clusters Database: \\ a progress report
}

   \subtitle{}

\author{
M. \,Castellani
        }

\offprints{M. Castellani}

\institute{
Istituto Nazionale di Astrofisica --
Osservatorio Astronomico di Roma, Via Frascati 33
I-00040 Monteporzio Catone (Roma)
\email{m.castellani@oa-roma.inaf.it}
}

\authorrunning{Castellani}

\titlerunning{Gclusters: a progress report}

\abstract{
The present status of Galactic Globular Clusters Database is briefly reviewed.
The features implemented at the time writing are described, as well as
plans for future improvements.
}

\maketitle{}

\section{Introduction}

The Galactic Globular Clusters Database
(briefly, {\it Gclusters}) \footnote{http://snipurl.com/gclusters}
is focused on presenting, in an organized way, a comprehensive list of bibliography,
parameters and data for each of the known globular cluster of the Milky Way (GGCs).
The need for a rational and organic assembly of these data is
well described in a famous paper (\citep{har96}):

{\it "The globular clusters in the Milky Way have proven thoughout this century
to be irrepleaceble objects in an amazingly wide range of astrophysical studies...
Year after year, it has proven important to have readily available
up-to-date lists of parameters for these unique objects."}

After more than ten years, these words appear even more true: the amount of
available data on GGCs has increased at a steady rate, following closely the enhanced
capability of the technical instrumentations. Not only we have new and more reliable
parameters for a great part of the known clusters, but - thanks to
modern surveys conducted in bands different from the visible, such 2MASS
(\citep{skr06}) -
several other objects keep going to increase our list of Milky Way clusters
(e.g.,\citep{fro07}, \citep{bon07}).
However, such data are inevitably
scattered among the various papers, so what is needed is a simple way
to have the relevant informations on a given cluster in a single source.

The Harris' catalogue of GGCs \footnote{http://snipurl.com/gclusters2}
is surely an unique resource for the
researchers, in what it provides an extensive list of parameters for all the
GGCs known at the time of its last revision (Feb. 2003): such a compilation is
accessible online and is composed by tables of parameters
available in form of flat text files.

\begin{figure}[]
\caption{
\footnotesize
 The main page of the Gclusters 
}
\end{figure}

Initially built around the Harris' compilation,
{\it Gclusters} is designed to allow a more flexible fruition of available data, so to make
possible things such as ordering clusters according to the value of a given parameter,
select objects whose parameters fall in a given interval, display related bibliography
and colour magnitude diagrams, or even drop a note pertinent to that cluster, to be displayed in
the website. Data are collected from a growing number of sources, such as NASA Astrophysics Data System (ADS),
Clement's variable stars \footnote{http://snipurl.com/gclusters3} pages, related websites, etc...


\section{A quick tour on {\it Gclusters}}

Documentation available for a given cluster is gathered in one page for the user's commodity.
For example, typing "M 3" (one of the most popular cluster in the database, according to the
access counts) in the search box, you obtain the output shown in Fig. 3

\begin{figure}[]
 \caption{
\footnotesize
 Listing "positional parameters" of the clusters (Gclusters' Table 1)
}
\end{figure}

 \begin{figure}[]
 \caption{
 \footnotesize
  The webpage for M3 
 }
 \end{figure}

In this page, you can browse the whole list of parameters available for the cluster,
together with a colour magnitude diagram
and a Digital Sky Survey image.
On the right
columns, you can find links to other related resources available on the web, as well as a direct
access to selected bibliography and NASA ADS search results.

The list of clusters are also conveniently divided into three tables that mimic the division
made by Harris (see Fig.2 for a partial view of Table 1), plus one table of "essential" bibliography.
Clusters in each of the three
tables can also be sorted according to the value of one of the listed parameters.

It is also possible to obtain list of clusters whose parameters match some given criteria.
Let's say that, in order to complete your (fundamental) paper,
you need to know (quickly) what are the globular clusters
that have metallicity greater than [Fe/H]=-1.6 and present a V magnitude
of {\it Horizontal Branch} less than 15.
Filling the {\it search page} with your data, you obtain a output a page with
the list of the eight clusters that match your requirements. A wide combination of searches
are possible.

\section{Technical Info \& Statistics}

This project is developed with Open Source software; specifically, it is a plane {\it "LAMP" application}
\footnote{http://snipurl.com/gclusters4}
In particular, storing the parameters
in form of tables of a {\it relational database}
instead of plane text files, makes possible to use them in a much more flexible way:
searches, ordered listings, and other queries on the data
can be performed easily from the Gclusters website. Adding new data and putting them in relation
with existing data is also a straightforward procedure.


For what concern the number of connections to the website,
taking as example a period starting from May 15 and ending to June 15, Gclusters website collected
a total of {\bf 3832 page views}, corresponding
to {\bf 1171 different visitors} (source: {\it Google Analytics}).
The day of maximum of page view is June 9, with a total of 493 visualized pages:
as inspection of the logs revealed, it was the effect of having been linked
from the NASA "Astronomical Picture Of the Day" (APOD), which on the same day
presented a nice image of the globular cluster M3
{\footnote{http://snipurl.com/gclusters5}.
The major number of visits for the quoted month came, in decreasing order,
from United States, Italy, Canada, Brazil and United Kingdom.

\section{Open to the scientific community}

It is possible to collaborate to the project at a wide range of "levels",
from pretty scientifical tasks (such as insertion of new data and bibliographic items)
to fairly technical ones (mainly HTML and PHP coding); anyway, even the disponibility
to simply "test" new pages could be precious for the developements of new features!
Note that the nature of the project make easy an Internet based collaboration:
people interested are warmly invited to contact me by Email.

\section{Future developments}

Admittedly, several branches of the project are still under developments.
Updated info  on the status of the project
can be found on the related blog
\footnote{http://globularclusters.wordpress.com}.
A list of possible improvement includes:
{\bf (a)}
   insertion in the database of addictional data compilations:
   for instance, dynamical data (e.g., \citep{agu88});
{\bf (b)}
   possibility to perform complex queries to select clusters that
   combine any range of given condition; possibility to refine searches;
{\bf (c)}
  availability of a  wider collection of CM diagrams and of a more complete bibliography.
{\bf (d)}
  insertion of data (periods, magnitudes...) for variable stars
  (e.g., \citep{cas03});
{\bf (e)} developement of (hopefully) {\it smart} procedures for
easy display and retrieval of relevant "row data" from main scientifical archive (e.g.,  ESO, HST...).
Recently, a connection was enstablished with people of the
WEBDA Open Cluster Database \footnote{http://www.univie.ac.at/webda/}, to explore the possibility
of realizing a common environment for both the databases,

\section*{A dedication}
{\it Gclusters} is dedicated to {\it Vittorio Castellani},
who passed away in May 2006.

\begin{acknowledgements}
The author thanks Luigi Pulone (INAF-OAR, Italy) for a number of stimulating discussions started before
the realization of {\it Gclusters}, and William E.Harris (McMaster University, Canada),
for useful remarks about an early release of the project.
\end{acknowledgements}

\bibliographystyle{aa}

\end{document}